\documentclass[aps,showkeys]{revtex4}

\usepackage{graphicx}
\usepackage[latin1]{inputenc}
\begin{document}

\title{EVALUATION OF PEAK-FITTING SOFTWARE FOR GAMMA SPECTRUM ANALYSIS}

\author{Guilherme S. Zahn}
\author{Frederico A. Genezini}
\author{Mauricio Moralles}
\affiliation{Instituto de Pesquisas Energéticas e Nucleares, Caixa Postal 11049, 05422-970, São Paulo, SP,
Brazil}

\begin{abstract}
In all applications of gamma-ray spectroscopy, one of the most important and delicate parts of the data analysis is the fitting of the gamma-ray spectra, where information as the number of counts, the position of the centroid and the width, for instance, are associated with each peak of each spectrum. There's a huge choice of computer programs that perform this type of analysis, and the most commonly used in routine work are the ones that automatically locate and fit the peaks; this fit can be made in several different ways -- the most common ways are to fit a Gaussian function to each peak or simply to integrate the area under the peak, but some software go far beyond and include several small corrections to the simple Gaussian peak function, in order to compensate for secondary effects. In this work several gamma-ray spectroscopy software are compared in the task of finding and fitting the gamma-ray peaks in spectra taken with standard sources of $^{137}$Cs, $^{60}$Co, $^{133}$Ba and $^{152}$Eu. The results show that all of the automatic software can be properly used in the task of finding and fitting peaks, with the exception of GammaVision; also, it was possible to verify that the automatic peak-fitting software did perform as well as -- and sometimes even better than -- a manual peak-fitting software.
\end{abstract}

\keywords{spectrum analysis; peak fitting; evaluation}

\maketitle

\section{Introduction}

In every quantitative application of gamma-ray spectrometry, one of the most important issues to be addressed is the identification in the spectra of the peaks associated with gamma-ray transitions and the precise determination of the position and area of each of these peaks. This process can be made manually, in a peak-by-peak basis, by a human operator; this approach, though seen as more trustworthy and safe by some researchers, is very tedious and time-consuming as, depending on the number of peaks in the spectrum, the analysis of a single spectrum can typically take 20--30 minutes or even more. On the other hand, there's a huge choice of computer software that can undertake the same job in a much shorter timeframe, saving time and manpower and thus allowing the analysis of more peaks in the same spectrum or of more samples in the same time; the question, though, is how well do these software perform in the tasks of identifying and determining the areas of the peaks, specially in a typical experimental spectrum, with something like 100 peaks that aren't always very well separated from each other and have areas that may differ in up to 3--4 orders of magnitude. This problem led IAEA to make an intercomparison on such software, which was published in 1997~\cite{1}, but since then new versions of some of those software were released and some new ones were introduced, so this work aims to update that comparison, as well as to make an explicit comparison of the automatic software to a manual peak fitting one.

\subsection{Identification of Peaks}
The first step an automatic gamma-ray spectrum analysis software must perform is the identification of the peaks in the spectrum; in this step, the two main tasks the software must undertake are: a) to identify peaks in the spectrum and differentiate them from simple fluctuations in the continuum background; and b) to identify and separate piled-up peaks.

\subsection{Peak Area Determination}
The simplest way to determine the area of a gamma-ray peak is simply to add up the counts from each of the channels in the peak range; the contribution of the continuum background is usually determined by averaging on one or two ``clean'' regions nearby and then subtracted from the result. In this approach, the uncertainty in the peak area assumed to be simply due to the statistical fluctuations in the areas determined, i.e., if the area of the peak is $A$, the full width of the peak (in channels) is $W_p$, the area of the background-only region is B and its width is $W_b$, then the net peak area ($N$) is:

\begin{equation}
N = A -- B \cdot \frac{W_p}{W_b}
\end{equation}

\noindent and, keeping in mind that $\sigma_A^2=A$ and $\sigma_B^2=B$, the uncertainty is given by:

\begin{equation}
\sigma_N = \sqrt{A + B \cdot \left(\frac{W_p}{W_b}\right)^2}
\label{eq:simplesum}
\end{equation}

There are some problems with this simple approach; when there are multiple peaks piled-up together, it is very difficult to properly determine the individual areas; in many cases the uncertainty obtained in this method is quite large and not very realistic; and the background contribution determined can be influenced by small, unidentified peaks, for instance.

The other usual way to calculate the peak areas consists in fitting some mathematical function -- usually a Gaussian (Eq.~\ref{eq:gaussian} -- to the peak; the background is then usually also fitted in a polynomial of the first or second degree. In this approach, a large part of the spectrum, with several peaks, is often fitted at once, thus allowing a much more precise determination of the real continuum background; this also allows a much more acute analysis of the parameters associated with the peak shapes, making the identification and separation of piled-up peaks much more precise~\cite{2}. The uncertainties in this type of calculation are determined much more realistic, as they are obtained directly from the function fits and, in many cases, should be smaller than the ones obtained in Eq.~\ref{eq:simplesum}. It should be noted, though, that in very weak or narrow peaks the peak is very badly formed and fitting its shape is difficult and can lead to unrealistic results; also, when the peaks are deformed by secondary effects as pile-up, incomplete charge collection and others, corrections must be added to the fitting function in order to avoid systematical errors.

\begin{equation}
Y(x) = A \cdot e^{-\frac{(x-\bar{x})^2}{2 \cdot s^2}}
\label{eq:gaussian}
\end{equation}

\section{Objective}
The objective of the present work is to compare different gamma spectrum analysis software in the task of identifying and fitting gamma-ray peaks in experimental spectra, in order to check if they render reliable and precise results, as needed for analytical spectrometry applications.

\section{Materials and Methods}
The experimental spectra used in this comparison were taken using standard sources of $^{133}$Ba, $^{152}$Eu, $^{60}$Co and $^{137}$Cs with activities ranging from 15 to 65 kBq in a 20\% efficiency germanium detector coupled to a DSP data acquisition system. The sources were counted for 3600s (live time) using the Genie 2000 software~\cite{3}, the source-detector distance was 3 cm and the dead times were never higher than 10\%.

The spectra were then analyzed using 5 automatic spectrum analysis software: Genie 2000~\cite{3}, GammaVision~\cite{4}, Hypermet PC~\cite{2}, as well as in-house developed software SAANI~\cite{5} and VISPECT. The analysis was performed using mostly the default settings; the only exceptions were the Hypermet PC software, where the ``Try'' setting was lowered to 2 (the default is 3) because it was fitting too many inexistent peaks, and the GammaVision software, where a manual FWHM calibration had to be made before it could properly fit the spectrum (with its default settings it would fit the peaks with too few channels, resulting in peak areas typically 10 times lower than the obtained using the other software).

In order to check for the possible disadvantages of automatic software, spectra were also analyzed using the manual peak fitting software IDeFix~\cite{6}, which allows for a very detailed and careful fit. As the version of Genie 2000 available only saves spectra in the proprietary format CNF, which can't be opened in any of the other software, the spectra were converted to CHN format using the Cambio software~\cite{7} -- there is another alternative for this conversion, IAEA's WinSPEDAC~\cite{8}, but this software showed problems when converting high-intensity peaks.

The comparison was then made first by fitting an efficiency calibration curve~\cite{9} to the data obtained with each software for the $^{133}$Ba and $^{152}$Eu sources, in order to check for the consistency of the fits. The efficiency function used is shown in Eq.~\ref{eq:eficiencia}, where the $P_i$ are the fitted parameters and $E_\gamma$ the gamma energy (in MeV), and the fit was performed using a Gauss-Marquardt non-linear least squares procedure implemented in the MATLAB environment.

\begin{equation}
\varepsilon \left(E_\gamma \right) = \left(P_1 \cdot e^{-P_2 \cdot E_\gamma} + P_3 \cdot e^{-P_4 \cdot E_\gamma} \right) \cdot e^{-P_5 \cdot \left(0.05757 \cdot E_\gamma^{-0.416}+0.000465 \cdot E_\gamma^{-2.943} \right)}
\label{eq:eficiencia}
\end{equation}

\section{Results and Discussion}
First of all, the ability of the software to identify peaks was checked using the data from the $^{152}$Eu source~\cite{10}, which has the most complex decay scheme, with many gamma transitions of different intensities. All software were able to identify the strongest peaks, with transition intensities (per 100 decays) ranging from 28.41 (121.8 keV) and 0.1437 (656.5 keV); the exception was GammaVision, which could not identify the 1089.7 keV transition ($I_\gamma=1.73$), which is overlapped with the stronger 1085.8 keV one  --  more on that will be discussed below. GammaVision could not identify much below that intensity, while most software identified properly and got peaks areas with uncertainties below 10\% (a decent threshold for spectrometry applications) for peaks with intensities down to 0.0738  (324.8 keV); Table~\ref{tab:found} shows a summary of these results, together with a comparison of the average of the relative uncertainties for each peak compared to the one obtained using the Genie 2000 software (chosen as a comparator as it was the one that identified more peaks).

\begin{table}[ht]
\begin{center}

\begin{tabular}{c|c|c}
Software	& Peaks Identified	& $\sigma_A/\sigma_A$(Genie 2000)	\\
\hline
Genie 2000 & 61 & 1 \\
GammaVision & 40 & 3.1 \\
HypermetPC & 54 & 1.8 \\
VISPECT & 60 & 1.05 \\
SAANI & 58 & 1.09 \\
IDeFix & 59 & 1.09 \\
\end{tabular}

\caption{Comparison of the ability to identify and fit peaks in the $^{152}$Eu spectrum.}
\end{center}
\label{tab:found}
\end{table}

The results of the chi-squares of fits of the efficiency calibration function are shown in Table~\ref{tab:efic}; for these fits, the peaks used were:

\begin{itemize}
\item For $^{133}$Ba~\cite{11}, all peaks except the 79 keV transition, which wasn't identified by most software;
\item For $^{152}$Eu~\cite{10}, all peaks identified by all software, except the 563, 756, 930, 990 and 1528 keV ones, where all software gave results clearly higher than expected, probably due to contamination of some sort (in the total, 56 peaks from the $^{152}$Eu source were used).
\end{itemize}

\begin{table}[ht]
\begin{center}

\begin{tabular}{c|c}
Software	& $\chi^2$ \\
\hline
Genie 2000 & 17.2 \\
GammaVision & 30.9 \\
HypermetPC & 9.1 \\
VISPECT & 10.5 \\
SAANI & 22.5 \\
IDeFix & 19.6 \\
\end{tabular}

\caption{Results of the chi-square of the fits of the $^{133}$Ba and $^{152}$Eu data to the efficiency calibration function.}
\end{center}
\label{tab:efic}
\end{table}

From these results it can be noticed that the GammaVision software was way off, yielding the highest uncertainties and, at the same time, the highest $\chi^2$ in the fit of the efficiency function, indicating that its results are imprecise and inconsistent. On the other hand, HypermetPC gave uncertainties that were, on average, more than 50\% larger that the ones given by Genie 2000, but that reflected in a much lower $\chi^2$ in the fit, indicating consistent results (in fact, HypermetPC's uncertainties were a lot larger for the most intense peaks, but mostly comparable to the others in the weaker ones); VISPECT did a very good job, too, reaching a $\chi^2$ comparable to that of HypermetPC, but with much lower uncertainties; SAANI and Genie 2000's performances were rather similar, both showing a rather large dispersion of the results, indicated by $\chi^2$ value considerably larger that the other software. It is interesting to note, though, that except for GammaVision, the results obtained by automatic software were compatible or better than the ones obtained using the manual fitting software IDeFix.

The last test was to compare the software in the task of fitting specific peaks. As each software led to a different efficiency calibration curve, the results were analyzed calculating the activity of the radioactive source using each specific transition and then comparing these results to the known activity of the source.

\begin{figure}[thb]
\begin{center}
\includegraphics[width=.4\textwidth]{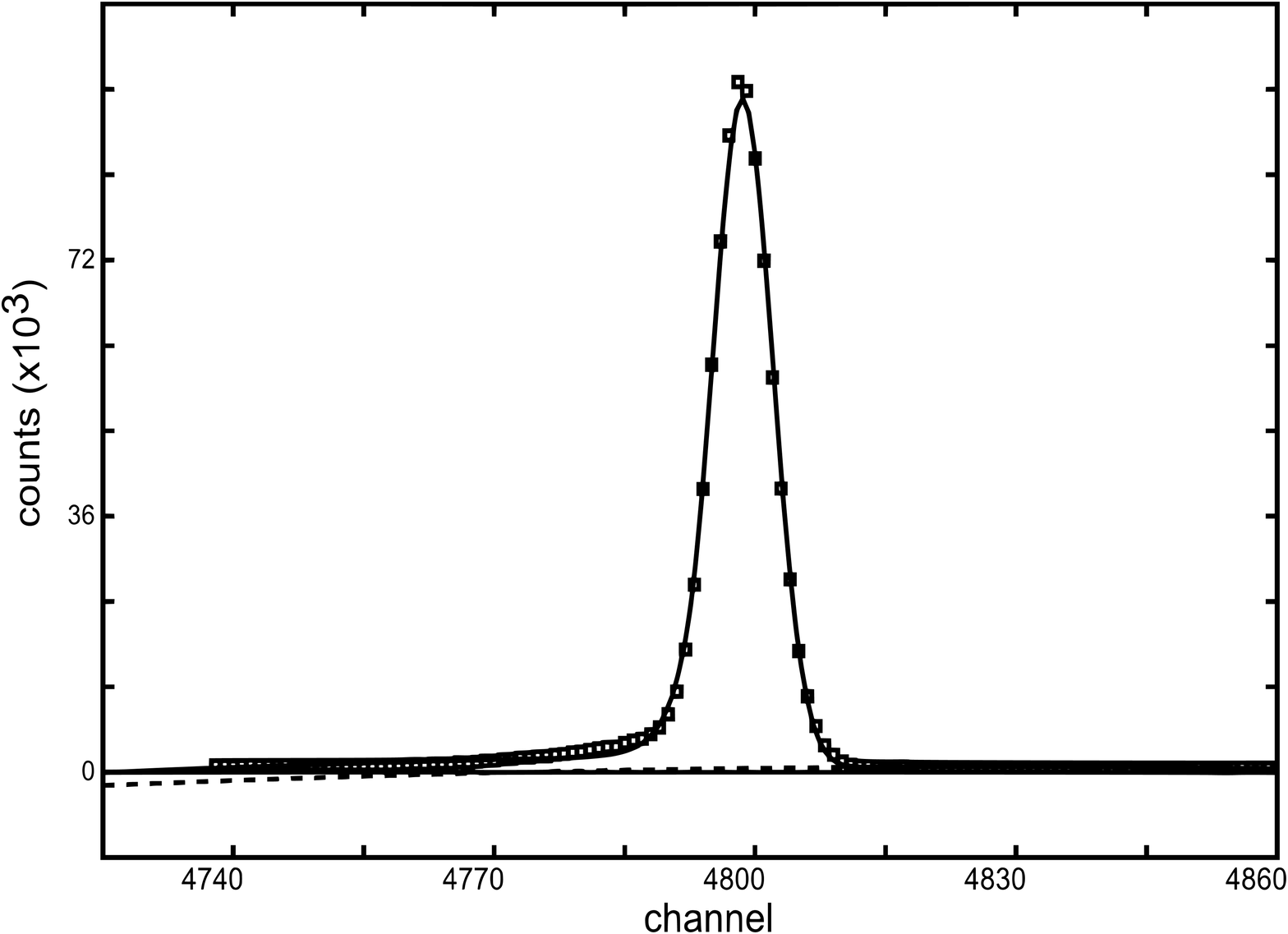}
\includegraphics[width=.4\textwidth]{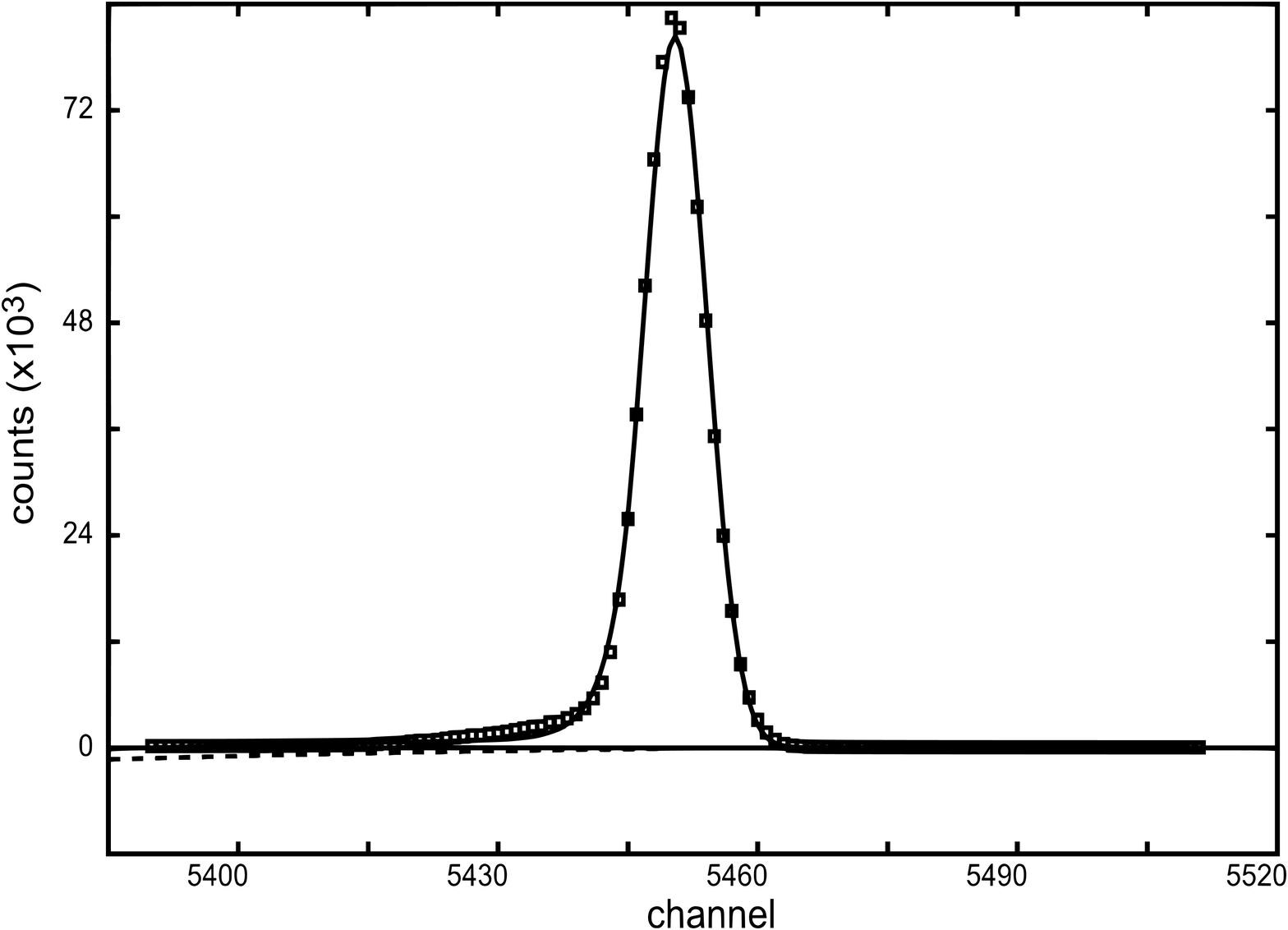}
\caption{Peaks at 1173 keV (left) and 1332 keV (right) from the $^{60}$Co source; the fit shown was made using the IDeFix software. \label{fig:Co}}
\end{center}
\end{figure}

The first test was to fit the two intense and well-separated gamma-ray transitions from $^{60}$Co~\cite{12}, with energies of 1173 and 1332 keV (Fig.~\ref{fig:Co}); the results, presented in Table~\ref{tab:Co}, show that all software did an excellent job in these simple transitions, although once again HypermetPC delivers larger uncertainties than the other software.

\begin{table}[ht]
\begin{center}

\begin{tabular}{c|c|c|c|c}
Software & Z(1173) & Z(1332) & $\sigma_{1173}$(\%) & $\sigma_{1332}$(\%) \\
\hline
Genie 2000 & -0.04 & 0.58 & 0.13 & 0.13 \\
GammaVision & -0.35 & -0.17 & 0.13 & 0.13 \\
HypermetPC & -0.091 & 0.10 & 0.83 & 1.20 \\
VISPECT & 0.22 & 0.63 & 0.10 & 0.10 \\
SAANI & 0.25 & 0.78 & 0.10 & 0.10 \\
IDeFix & 0.37 & 1.30 & 0.12 & 0.12 \\
\end{tabular}

\caption{Comparison of the activity determined for the $^{60}$Co source using each software; Z(1173) and Z(1332) are the Z-Scores calculated using the 1173 and 1332keV transitions, respectively, and $\sigma_{1173}$(\%) and $\sigma_{1332}$(\%) are the area percent uncertainties.}
\end{center}
\label{tab:Co}
\end{table}

\begin{figure}[thb]
\begin{center}
\includegraphics[width=.4\textwidth]{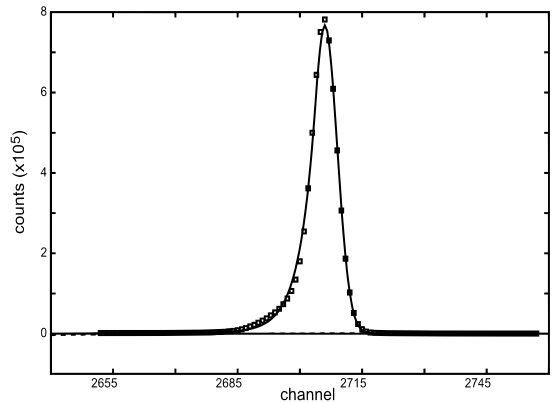}
\caption{Peak at 663 keV from the $^{137}$Cs source; the fit shown was made using the IDeFix software. \label{fig:Cs}}
\end{center}
\end{figure}

A tougher test was the fit of the 662 keV transition from $^{137}$Cs~\cite{13}; this transition is too intense and often show some deviation from the regular Gaussian shape (Fig.~\ref{fig:Cs}); the results (Table~\ref{tab:Cs}) show that most software underestimated the activity of the source; the exceptions were the HypermetPC (which would have gotten the activity perfectly right even without the larger uncertainties) and the IDeFix softwares -- in this case, probably, the more complex fitting procedures in these software accounted for the better results. Genie 2000 and SAANI also led to decent -- albeit not perfect -- results, and VISPECT and GammaVision underestimated heavily the peak's area.

\begin{table}[ht]
\begin{center}

\begin{tabular}{c|c|c}
Software & Z & $\sigma_{662}$(\%) \\
\hline
Genie 2000 & -1.49 & 0.045 \\
GammaVision & -3.24 & 0.051 \\
HypermetPC & -0.01 & 0.96 \\
VISPECT & -4.16 & 0* \\
SAANI & -2.07 & 0* \\
IDeFix & 0.39 & 0.049 \\
\end{tabular}

\caption{Comparison of the activity determined for the $^{137}$Cs source using each software; Z are the Z-Scores and $\sigma_{662}$(\%) are the area percent uncertainties;\newline
* -- These software only report uncertainties larger than 0.1\%.}
\end{center}
\label{tab:Cs}
\end{table}

\begin{figure}[thb]
\begin{center}
\includegraphics[width=.4\textwidth]{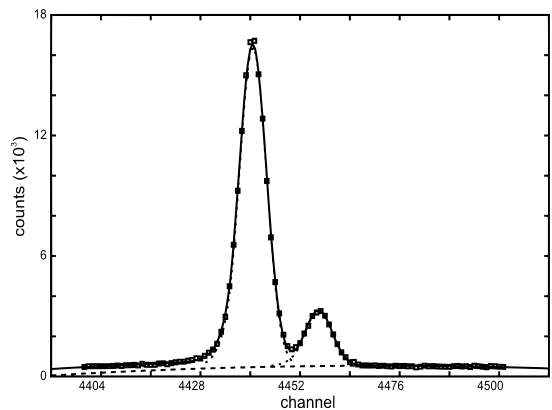}
\caption{Doublet peak at 1086 and 1090 keV from the $^{152}$Eu source; the fit shown was made using the IDeFix software. \label{fig:Eu}}
\end{center}
\end{figure}

To check for the ability to properly separate piled-up peaks, the software were compared in the analysis of the 1086--1090 keV doublet in $^{152}$Eu (Fig.~\ref{fig:Eu}); the results, presented in Table~\ref{tab:Eu}, show that all the software except the GammaVision were able to separate the two peaks; nevertheless, Genie 2000 and IDeFix were the only ones to get the intensity of the 1090 keV peak correctly -- once again HypermetPC gave larger uncertainties, specially on the stronger peak.

\begin{table}[ht]
\begin{center}

\begin{tabular}{c|c|c|c|c}
Software & Z(1086) & Z(1090) & $\sigma{1086}$(\%) & $\sigma_{1090}$(\%) \\
\hline
Genie 2000 & -0.76 & 0.005 & 0.29 & 0.79 \\
GammaVision & 7.59 & * & 0.40 & * \\
HypermetPC & 0.44 & 3.08 & 2.85 & 1.49 \\
VISPECT & -0.30 & -1.96 & 0.30 & 1.00 \\
SAANI & -0.38 & -3.76 & 0** & 1.10 \\
IDeFix & -0.31 & 0.52 & 0.40 & 0.89 \\
\end{tabular}

\caption{Comparison of the activity determined for the $^{152}$Eu source using each software; Z(1086) and Z(1090) are the Z-Scores calculated using the 1086 and 1090 keV transitions, respectively, and $\sigma_{1086}$(\%) and $\sigma_{1090}$(\%) are the area percent uncertainties;\newline
* The GammaVision software was unable to separate the two peaks.\newline
** These software only report uncertainties larger than 0.1\%.}
\end{center}
\label{tab:Eu}
\end{table}

\begin{figure}[thb]
\begin{center}
\includegraphics[width=.4\textwidth]{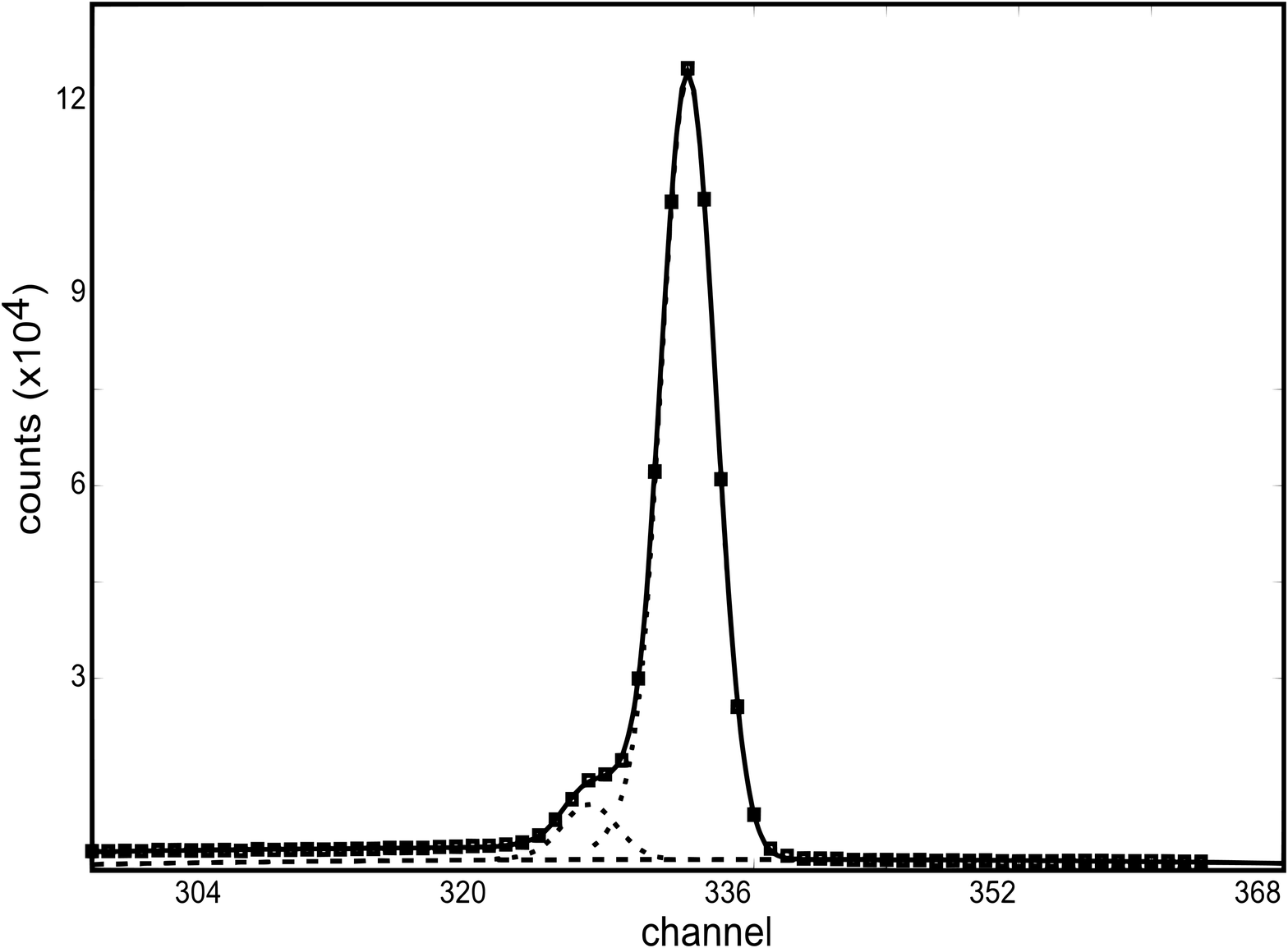}
\caption{Doublet peaks at 80 and 81 keV from the $^{133}$Ba source; the fit shown was made using the IDeFix software. \label{fig:Ba}}
\end{center}
\end{figure}

The last test was to separate the very close doublet at 80 and 81 keV from the $^{133}$Ba source (Fig.~\ref{fig:Ba}). In this case the only automatic software that was able to identify both transitions was HypermetPC; the results of each software are presented in Table~\ref{tab:Ba}, where it can be seen that once again HypermetPC's results showed the largest uncertainties.

\begin{table}[!ht]
\begin{center}
\begin{tabular}{c|c|c|c|c}
Software & Z(80) & Z(81) & $\sigma{80}$(\%) & $\sigma_{81}$(\%) \\
\hline
Genie 2000 & * & -0.03 & * & 0.15 \\
GammaVision & * & -0.10 & * & 0.17 \\
HypermetPC & -0.91 & 0.27 & 5.17 & 0.44 \\
VISPECT & * & -0.11 & * & 0.20 \\
SAANI & * & -0.26 & * & 0.20 \\
IDeFix & -1.03 & -0.32 & 1.40 & 0.18 \\
\end{tabular}

\caption{Comparison of the activity determined for the $^{133}$Ba source using each software; Z(80) and Z(81) are the Z-Scores calculated using the 80 and 81 keV transitions, respectively, and $\sigma_{80}$(\%) and  $\sigma_{81}$(\%) are the area percent uncertainties;\newline
* The software was unable to separate the two peaks.}
\end{center}
\label{tab:Ba}
\end{table}

\section{Conclusions}
The tests performed in this work showed that most automatic gamma spectrometry software deliver good and reliable results; the only exception was GammaVision, which failed to find many important peaks and often showed results that were a lot worse than any of the other software -- a previous version of this software, tested in the IAEA intercomparison~\cite{1}, also showed some problems in the peak area determination. The other software tested all seemed to meet the minimum requirements for daily routine; HypermetPC was the best in finding weak peaks and the only one to correctly identify the weak 80 keV transition in $^{133}$Ba, possibly due to an internal peak library, but often delivered larger uncertainties, especially in the stronger peaks; Genie 2000 performed very well all around and seems like a nice choice; SAANI and VISPECT both did a good job in most cases, but failed in others. Also, with the exception of GammaVision, the automatic software performed as well as or even better than the manual peak fitting software IDeFix consuming only a small fraction of the time it takes for a manual analysis, indicating that it is indeed more interesting to use them for daily work.

\end{document}